\documentclass{svproc}
\usepackage{cite}
\usepackage{url}  % This makes \url work
\PassOptionsToPackage{hyphens}{url}
\usepackage[hidelinks]{hyperref}
\hypersetup{breaklinks=true}
\usepackage{multicol}

\usepackage{enumitem}
\usepackage{lipsum}
\usepackage[dvipsnames]{xcolor}

\newcommand{\refsec}[1]{Section~\ref{#1}}
\newcommand{\reffig}[1]{Figure~\ref{#1}}
\newcommand{\reftab}[1]{Table~\ref{#1}}

\newcommand{\etal}{~et~al.}
\newcommand{\ttt}[1]{{\small \texttt{#1}}}

\expandafter\def\expandafter\UrlBreaks\expandafter{\UrlBreaks% save the current one
  \do\a\do\b\do\c\do\d\do\e\do\f\do\g\do\h\do\i\do\j%
  \do\k\do\l\do\m\do\n\do\o\do\p\do\q\do\r\do\s\do\t%
  \do\u\do\v\do\w\do\x\do\y\do\z\do\A\do\B\do\C\do\D%
  \do\E\do\F\do\G\do\H\do\I\do\J\do\K\do\L\do\M\do\N%
  \do\O\do\P\do\Q\do\R\do\S\do\T\do\U\do\V\do\W\do\X%
  \do\Y\do\Z\do\*\do\-\do\~\do\'\do\"\do\-}%

\usepackage{wrapfig}
\usepackage{subcaption}
\usepackage{graphicx}
\graphicspath{{images/}}
\usepackage{tikz}

\usepackage[export]{adjustbox}

\let\oldsubsec\subsection
\renewcommand{\subsection}[1]{\vspace{-10pt}\oldsubsec{#1}\vspace{-5pt}}
\let\oldsubsubsec\subsubsection
\renewcommand{\subsubsection}[1]{\vspace{-10pt}\oldsubsubsec{#1}\vspace{-5pt}}
\let\oldpar\paragraph
\renewcommand{\paragraph}[1]{\vspace{-10pt}\oldpar{#1}}

\begin{document}
\mainmatter              % start of a contribution
\title{Beartooth Relay Protocol: Supporting Real-Time Application Streams over LoRa}
\titlerunning{Beartooth Radio Protocol}  % abbreviated title (for running head)
\author{Batuhan Mekiker\inst{1} \and Mike Wittie\inst{1} \and Jefferson Jones \and Michael Monaghan\inst{2}}
\authorrunning{Batuhan Mekiker et al.} % abbreviated author list (for running head)
% \tocauthor{Ivar Ekeland, Roger Temam, Jeffrey Dean, David Grove,
% Craig Chambers, Kim B. Bruce, and Elisa Bertino}
%
\institute{Montana State University, Bozeman, MT, USA
\and
Beartooth Radio Inc., Bozeman, MT, USA}

\maketitle              % typeset the title of the contribution

\begin{abstract}
The near-ubiquitous availability of wireless connectivity lets users take advantage of a large variety of mobile applications.
This connectivity predominantly comes as cellular and WiFi, limiting users to available infrastructure.
At the same time, commercial efforts for infrastructure-less connectivity do not support mobile application traffic.
In this paper, we present a new LoRa radio and a relay protocol capable of supporting real-time application traffic on point-to-point and multihop connection.
Our solution has the potential to extend mobile application functionality beyond infrastructure coverage areas.

\end{abstract}

\vspace{-10pt}
\section{Introduction}
\vspace{-5pt}

%% SOCIETAL PROBLEM
%% People want ubiquitous and performant networks.

People have grown used to wireless internet connectivity.
They expect to keep up with news, work email, and stay in-touch through a variety of messaging and video apps.
Indeed, essential functions of daily life such as transportation, shopping, lodging, and payments rely on mobile applications communicating with cloud backends and, soon, edge computing servers in near real-time.
Unfortunately network coverage, or adequate provisioning, is not ubiquitous and areas such as rural roadways and developing regions need means to bridge gaps between cellular and WiFi coverage.  
% As one of the authors can attest, travelling through Europe with limited cellular connectivity and no access to mobile apps is no picnic. 

%% TECHNICAL PROBLEM
%% Existing networks have limited coverage and require a lot of power.

Modern connectivity comes predominantly through two types of technology: cellular and WiFi.
While both have evolved over the years to offer impressive throughput and even rapidly decreasing latency~\cite{paper:NETW_EVOL}.
Networks based on these technologies remain reliant on infrastructure which limits their availability and performance outside of major cities~\cite{paper:FCC_REP}.
At  the  same  time, infrastructure-less connectivity solutions are not well-suited to fill the gap for mobile applications.
Walkie-talkie and amateur~(HAM) radios do not offer data services integrated with cellphones primary due to regulatory limitations precluding encryption~\cite{paper:FCC_Title47}. 
Satellite communications, while available on some smartphones, remain expensive and performance-limited, though recently planned new constellations may change that~\cite{paper:STARLINK}. %, paper:STARLINK2}.
In the industrial, scientific and medical~(ISM) band solutions such as DASH7, Z-Wave, and Sigfox target sensor networks and provide energy-efficient, but low data-rate connectivity~\cite{paper:DASH7, paper:LPWAN_DATA_RATE}.
802.11ah provides higher data rates, but at a more restricted range~\cite{paper:80211ah}.
Recently introduced Long Range~(LoRA) networks can cover large area due to the long range permitted by the chirp spread-spectrum~(CSS) modulation, but lack support for mobile application traffic~\cite{paper:LoRA_Study}.
% and multiple coordinated gateways running the LoRaWAN protocol~\cite{paper:LoRA_Study}.
% On the other hand, 
Finally, GoTenna offers a Frequency Shift Keying~(FSK) ISM network with a flooding protocol focused on short message delivery, but lacks throughput and latency to support mobile applications.

%% TECHNICAL SOLUTION
%% Beartooth offers a third leg of connectivity.

To bridge the gap between coverage limitations of cellular and WiFi networks and the lack of support for mobile applications in the ISM band, we offer the following contributions:

\vspace{-15pt}

\begin{enumerate}%[itemindent=2em, leftmargin=1em, rightmargin=1em]
\item 
We describe a new Beartooth LoRa radio designed to address the limitations of cellular and WiFi coverage and low data rate of existing infrastructure-less networks.
Beartooth's goal is to support throughput and latency of existing mobile application traffic.
Indeed, Beartooth is the first LoRa system to deliver real-time voice on point-to-point connections.

\item 
We present a new scheduled multihop protocol, the Beartooth Relay Protocol~(BRP), to extend the system's support for mobile application traffic beyond point-to-point connections.
While multihop connectivity can connect a Beartooth network to cloud servers that support mobile application functionality, we believe that many mobile applications could be adapted to provide much of their functionality with local connectivity only.

\item 
We experiment with a smartphone-based deployment of BRP and report on the feasibility and limitation of software-based link-layer implementations.
Placing link-layer logic on the smartphone reduces Beartooth hardware costs, provide the flexibility for faster experimentation, and simplify protocol updates.
On the other hand, a software implementation carries with it the inefficiencies of cross-layer communications.
% While placing link-layer logic on the smartphone could be cost effective and offer greater flexibility than placing it on the radio hardware, it carries with it inefficiencies of cross-layer communications.
We demonstrate that running the link-layer in software is possible and provides enough throughput to carry voice traffic, although with additional latency.

% We demonstrate that in spite of these, a software implementation of BRP provides enough throughput to carry voice traffic, although with additional latency.
% Based on our experimentation with software-based BRP we identified several improvements we plan to make as we transition to an on-chip BRP implementation.
% We also discuss protocol tradeoffs and improvements to meet latency requirements in future implementations of the protocol.

\item
Finally, we present a detailed measurement study of Beartooth radios running BRP.
We quantify the overhead of software protocol implementation as well as its performance in terms of latency, throughput, and battery power draw.
We extend these results with a discussion on protocol limitations and improvements, including plans for on-chip implementation.
\end{enumerate}

\vspace{-15pt}

%% EXPECTED IMPACT
%% Beartooth enables a third leg of communication.

Our results show that Beartooth networks have the potential to form a \emph{third leg} of consumer connectivity to fill in the coverage gaps between cellular and WiFi networks and supporting mobile application traffic.
The adoption of Beartooth radios in smartphones, already underway in the Sonim~XP8, paves the way to ubiquitous and constant connectivity beyond infrastructure-based wireless networks.

%% ORGANIZATIONAL PARAGRAPH

The rest of this paper is organized as follows.
In \refsec{sec:related_work} we discuss the limitations of existing communication technologies.
\refsec{sec:radio} introduces the Beartooth radio and \refsec{sec:protocol} its multihop protocol suite.
In \refsec{sec:evaluation} we present a measurement study of the performance within a Beartooth network.
\refsec{sec:lims_and_fw} discusses limitations of our approach and an outline of future work.
Finally, we conclude in \refsec{sec:conclusions}.

\vspace{-10pt}
\section{Related Work}
\vspace{-5pt}
\label{sec:related_work}

To frame the need for an alternative to the cellular/WiFi duopoly, we briefly discuss the limitations of these networks.
We then explore the work on infrastructure-less communications within the cellular, WiFi, and ISM bands.

\subsection{Limitations of Cellular and WiFi Networks}

The 2019 Federal Communications Commission~(FCC) mobile coverage report states that 
% 1\% of Americans in rural areas and 3\% of Americans on Tribal Lands still lack cellular coverage~\cite{paper:FCC_REP}. 
31\% Americans in rural areas lack access to 25\,Mbps/3\,Mbps LTE speeds and 27\% lack access to terrestrial internet~\cite{paper:FCC_REP}.
Even in areas nominally covered by cellular service, the economies of tower deployments create frequent occlusion zones, especially in mountainous areas, due to shadowing at longer distances~\cite{paper:WIFI_PYATTAEV}.
% Our analysis of Montana's coverage map on RootMetrics shows holes in cellular coverage along many roadways~\cite{paper:rootmetrics_mt}.
At the same time it is crucial to provide full coverage, because network connectivity is increasingly important for safety-critical applications, including self-driving vehicles~\cite{paper:MC_reliability}. % paper:GSMA_MC

% In spite of FCC's efforts to promote coverage, the goal of full coverage comes up against the economic viability of infrastructure deployment in low population-density areas as well as regulatory barriers to invest in certain areas~\cite{paper:FCC_REP}.
% While satellite connectivity has the promise of filling this gap, network performance and data costs remain too high for most applications and varies around \$0.65 - \$1 per minute of internet connection depending on the underlying technology with bandwidth of 2.4\,Kbps~\cite{paper:satellite, paper:Sat_cost}.

\subsection{Promise of Infrastructure-less Networks}

Many mobile applications may be able to perform their functions with temporarily limited or no access to back-haul networks.
Accordingly, there have been several efforts to create device-to-device~(D2D) links between mobile devices.

\subsubsection{D2D in cellular}

Doppler\etal\ put forward one of the first D2D proposals for cellular networks~\cite{paper:LTE_Doppler}.
The cellular tower manages direct connections between user equipment~(UE), which achieve lower latency and greater spectral efficiency than by forwarding their traffic through the evolved NodeBs~(eNBs), or base stations.
% To limit interference, UE restricts its power and thus communication range.
%
% Lei\etal\ describe the reluctance of cellular operators to support D2D due to cost and threats to profitability~\cite{paper:LTE_leilei}.
% While device makers are generally interested in D2D, the proposed approaches still rely on operator cooperation~\cite{paper:Nokia, paper:Ericsson, paper:Intel}.
%
The evolution of D2D in cellular networks is LTE~Direct, defined in 3GPP Release 12, which enables discovery and connection establishment among cellular devices without eNB involvement.
LTE~Direct range, however, is limited to 500 meters due to low antenna height, which is insufficient for robust connectivity, or multi-hop forwarding, in many rural areas.
% Lin\etal\ evaluate an experimental LTE~Direct deployment and demonstrate loss of throughput due to lack of coordination among devices that might make operators loathe to adopt it~\cite{paper:LTE_LIN_ANDREWS}.
% They conclude that operators might be reluctant to relinquish the control over their spectrum to UEs and that in practice LTE~Direct would require a significant network re-engineering effort.
% For a discussion of D2D approaches in cellular networks see a survey by Mach\etal~\cite{paper:D2D_SURVEY}. 
\looseness-1

\subsubsection{D2D in WiFi}

WiFi networks can establish D2D communications through WiFi Direct~\cite{paper:WIFI_CAMPS_MUR}.
The approach borrows from the existing 802.11 protocols and allows devices to negotiate which of them will act as an access point~(AP).
% As such, this approach limits the range of the networks to two hops.
% Because of the reliance on 802.11 security mechanism, group establishment may take as much as 8\,s which precludes the use of WiFi Direct in high mobility scenarios with frequently changing network topology and group membership.
Lei\etal\ have evaluated the range in WiFi Direct networks and found it limited to 656 inches~\cite{paper:LTE_leilei}.
Pyattaev\etal\ have also found that WiFi Direct connects suffer from interference on the unlicensed band as well as sharp increases in latency as the amount of traffic offloaded cellular traffic increases~\cite{paper:WIFI_PYATTAEV}.

\subsubsection{D2D in ISM}

In spite of its initial intent for device communications, the ISM band has increasingly been used for general telecommunications.
Solutions such as 
Sigfox~\cite{paper:UNB_Overview}, 
Z-Wave~\cite{paper:zwave},
802.11ah (WiFi HaLow)~\cite{paper:80211ah}, 
and 802.15.4 low-rate wireless personal area networks (LR-WAN)~\cite{paper:802154}
support the sporadic exchange of short messages between devices do not meet the QoS requirements of mobile application traffic.
DASH7 supports star topologies with data rates of 13\,kbps/16 channels, 50/8, and 166/4~\cite{paper:DASH7}.
% Relayed D2D communications need to be configured by the network administrator during deployment~\cite{paper:DASH7}.
Experiments show practical range of 900\,m~\cite{paper:dash7_range1}, which give DASH7 limited potential to patch coverage in rural scenarios. %paper:dash7_range2
% While the advertised range is 3\,km experiments with the current DASH7~1.1 specification show that the lowest rate modulation establishes reliable links at no more than 900\,m~\cite{paper:dash7_range1, paper:dash7_range2}.
% As such, the application of DASH7 to patch coverage in rural scenarios is limited.
GoTenna adopts FSK and a proprietary Aspen Grove dissemination protocol to relay text messages and other low-rate data over multiple hops albeit with no guarantees on delay. % paper:gotenna_eval}. %paper:gotenna_about
% Because Aspen Grove relies on opportunistic forwarding GoTenna does not guarantee bounds on delay which reaches into seconds.
% The GoTenna Mesh device has 1\,mi range~\cite{paper:gotenna_white_paper}, while the more recent GoTenna Pro, has an advertised range of 4\,mi~\cite{paper:gotenna_pro}.

Semtech introduced LoRa chirp spread spectrum~(CSS) chip in 2009.
The main advantage of LoRa is its range -- radios based on Semtech chips support data rates between 0.3 and 11\,Kbps and robust links at distances up to 9\,km in urban and over 30\,km in rural scenarios~\cite{paper:LoRaWan_perf, paper:LoRaWan_perf2}.
The LoRa Alliance publishes a carrier sensing multiple access~(CSMA) LoRaWAN protocol, which allows devices to communicate via gateways~\cite{paper:lorawan}.
However, LoRaWAN error rates of 50\% on low data rate channels make it unsuitable for application streams with QoS requirements~\cite{paper:symphony}.
% Closest to our approach are Symphony Link and GoTenna~\cite{paper:gotenna, paper:symphony}.
Symphony Link uses a time division multiple access~(TDMA) protocol across multiple frequencies to schedule LoRa transmissions between its nodes and a gateway, but with the low rate of 37\,B packets every 2 seconds~\cite{paper:symphony}.
% While an improvement over LoRaWAN in terms of reliability and support for large files, such as firmware updates, Symphony Link is still intended for sensor networks with the highest level of node throughput at 4 times 37\,B packets every 2 seconds.

Recent research proposes new approaches to LoRA-based messaging.
Cardenas\etal\ propose a messaging system on top of LoRaWAN for emergency communications~\cite{paper:cardenas_msging}.
Hochst\etal\ propose to address rural connectivity gaps with a custom System-on-a-Chip(Soc) LoRa transceiver~\cite{paper:hochst_crisis_com}.
Mai\etal\ propose multi-hop, schedule communication over LoRa for low latency IoT applications~\cite{paper:mai_multihop_lora}.
Lundell\etal\ on the other hand, propose a mesh networking protocol based on Ad-hoc On-Demand Distance Vector (AODV) and Hybrid Wireless Mesh Protocol (HWMP)~\cite{paper:lundell_lora_mesh}.
Leonardi\etal\ underline the need for bounded end-to-end delay in IoT applications and introduce {RT-LoRa} to scheduling for real-time LoRa traffic~\cite{paper:leonardi_rt_lora}.
However, none of the proposed solutions provide enough link bandwidth for voice flows.

\vspace{5pt}

\noindent
In summary, there is a need for a third-leg of mobile communications that fills the coverage gaps left by cellular and WiFi approaches, while providing communication channels able to carry the traffic of modern mobile applications.

% \todo{mwittie 7/24/19: look into , LTE-M, and Weightless}

% \begin{figure}
%     \centering
    
%     \includegraphics[width=\linewidth]{images/domain_diag.png}
%     \caption{Figure representing the domain of existing technologies and the gap in the domain along with the representation of proposed solution \cite{paper:LPWAN_FIG3}}
%     \label{fig:my_label}
% \end{figure}

% \todo{mwittie 7/30/19: discuss symphony link \url{http://info.link-labs.com/hubfs/LPWAN_Technology_Explained.pdf?t=1509723653738&utm_source=hs_automation&utm_medium=email&utm_content=39440268&_hsenc=p2ANqtz-8bp06MWAgB4DWiWEM6GxmWqZ53Au8tyJThPT2AF}} 
% \cite{paper:symphony}

% Bullet points:
% - Although LoRa is a great MAC layer, link layer application for LPWAN LoRaWAN has lacking functionality.

% - with lorawan PER is above \%50 and mission critical application don't allow this high of PER

% - Easy to maintain with over the air updates. Technically  it's possible for Lorawan as well but the unreliability doesn't allow that.

% -Different type of protocol, Listen before talk, and freq. hopping allow surpassing duty cycle limit

% -Allow repeaters to gain access to larger areas.

% -Implemented QoS for delay-sensitive packets.

% -network configuration done by configuring gateway only. (like our config id)

% -adaptable transmit power and SF for balanced link budget. Closer nodes uses less power to transmit and lower SF (faster data rate), while further ones use higher transmit power with higher SF (slower transmission).

% - symphony has 4 times the capacity of lorawan. with scheduling, coordination and  managing resources.

% - cost of using lorawan \$20k/year, whereas symphony has no membership fees.

\vspace{-10pt}
\section{Beartooth Radio}
\vspace{-5pt}
\label{sec:radio}

\begin{wrapfigure}{r}{0.4\linewidth}
    \centering
    \vspace{-20pt}
    \begin{subfigure}{0.4\textwidth}
         \centering
         \includegraphics[width=\textwidth]{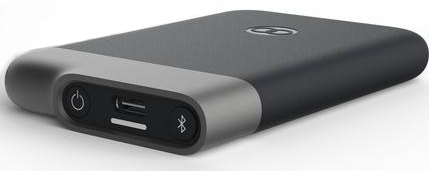}
         \caption{Beartooth consumer device.}
         \label{fig:radio_packaging}
     \end{subfigure}
     
     \vspace{5pt}
     \begin{subfigure}{0.4\textwidth}
         \centering
         \includegraphics[width=\textwidth]{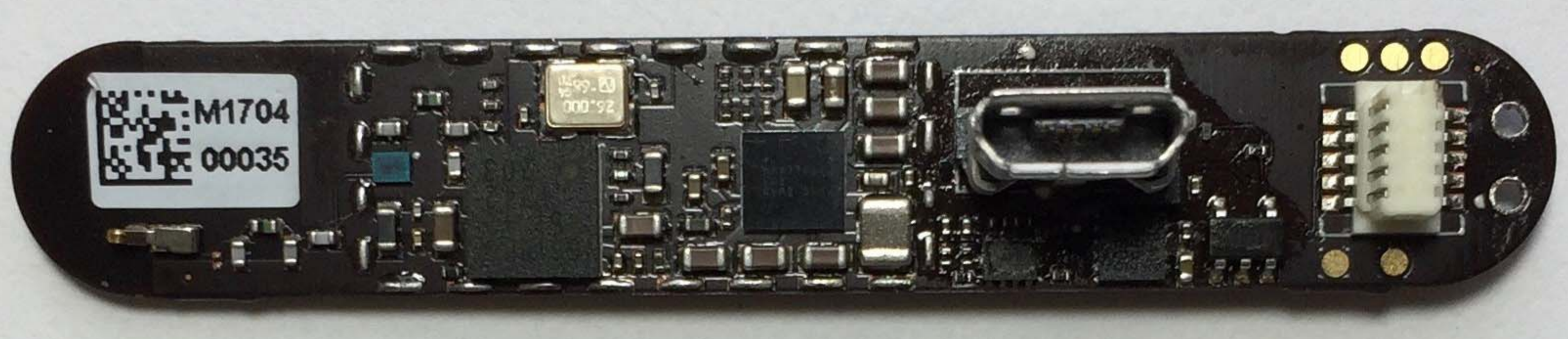}
         \caption{Beartooth radio board front.}
         \label{fig:board_front}
     \end{subfigure}
     
     \vspace{5pt}
      \begin{subfigure}{0.4\textwidth}
         \centering
         \includegraphics[width=\textwidth]{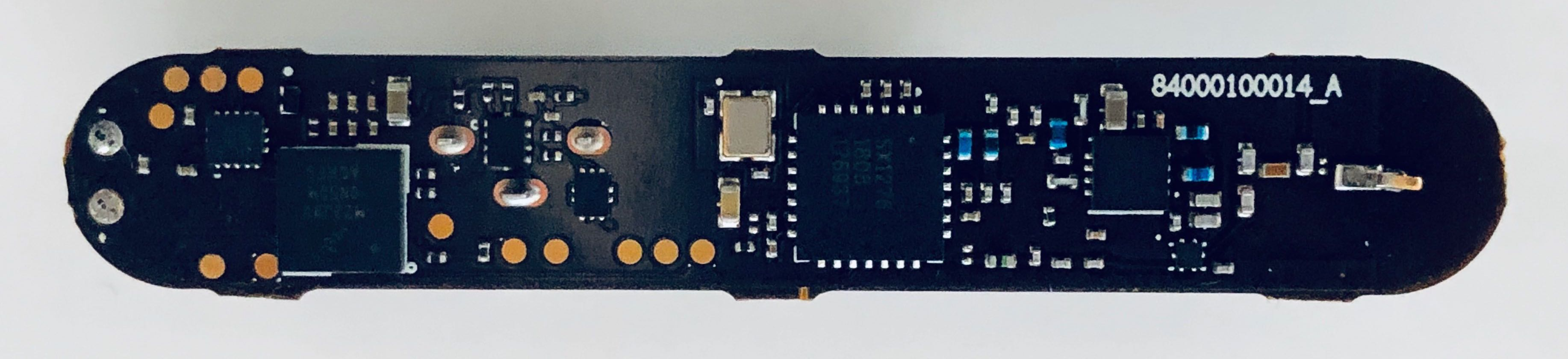}
         \caption{Beartooth radio board back.}
         \label{fig:board_back}
     \end{subfigure}
     \vspace{-5pt}
    \caption{Beartooth radio.}
    \label{fig:radio}
    \vspace{-25pt}
\end{wrapfigure}

The Beartooth radio, shown in \reffig{fig:radio}, is a pocket consumer device that connects to smartphones through Bluetooth and establishes a multihop network with other Beartooth radios through LoRa.
The radio is based on the Semtech SX1276 chipset coupled with an 11\,cm dipole antenna.
% The electronic package is much smaller than, for example, GoTenna's second generation radio~\cite{paper:gotenna_teardown}.

The SX1276 chipset modulates a chirp spread-spectrum~(CSS) radio signal in the 900\,MHz band amplified to 30\,dbm at the antenna.
LoRa encodes information with chirps, or transmissions of rising frequencies within the width of a channel (125\,kHz or 250\,kHz), where the starting frequency of chirp indicates a symbol~\cite{paper:LoRA_Facts}.
The slope of a chirp is a function of channel width and the spreading factor, which defines the duration of each chirp and its resiliency to radio interference.
% -- the wider the channel and the longer the chirp the more resilient the transmission to interference as these make it easier for the receiver to determine the starting frequency of the chirp.
The SX1276 provides seven spreading factors, SF6 to SF12, while SF6 provides the highest data rate, 37.5\,Kbps, SF12 provides the greatest robustness and therefore range with 260\,bps \cite{paper:LoRA_Study}.
The CSS encoding gives LoRa resilience to multipath effects, fading, and Doppler frequency shifts~\cite{paper:LoRA_Mod}.
LoRa also protects bits in transmission with a configurable error correction rates using Hamming codes and with a cyclic redundancy check~(CRC).
Finally, the Beartooth radio achieves collision resistance by frequency hopping between channels on successive symbols.
A pair of communicating nodes selects one of 50 orthogonal hopping sequences~\cite{paper:LoRA_Mod}.
The current, proprietary implementation of a point-to-point Beartooth protocol makes this the first LoRa system to support real-time voice communications (\ttt{https://youtu.be/ECZRpeSu-EM}).% ~\cite{paper:rt_voice}.
% \footnote{Recorded demo available at \url{https://youtu.be/ECZRpeSu-EM}}

% For comparison with other consumer electronics, the cost of the Beartooth radio is \$125.
% The technology is also available as a Beartooth System on a Module~(BSOM), which may be integrated with other hardware, such as the Sonim~XP8 cellphone~\cite{paper:XP8}, at a cost of \$6-\$8 depending on the volume.
% As such, Beartooth radios have the potential to be deployed alongside cellular and WiFi chips on modern cellphones and in other consumer devices.

\vspace{-10pt}
\section{Beartooth Relay Protocol (BRP)}
\label{sec:protocol}
\vspace{5pt}

% \todo{Discus what happens if control frames get lost - talk about our loss rate a bit}
% \todo{Improve the explanation of the figures - need to explain the timing of the different timeslots as well as packet lengths/contents}
% \todo{the evaluation of the achievable cycle time should have been done based on a
% mathematical model considering the packet size, number of nodes, guard time, etc,
% as explained above}
% \todo{discuss scheduling and how we assign data to voice rather than data streams}
% \todo{discuss addressing of packets - explicit vs implicit by timeslot}
% \todo{what is the duration of a cycle, how do we calculate it?}
% \todo{what happens if a node receives RLY\_ANNC from several relays?}
% \todo{how does the protocol deal with duty cycle limitations?}
% \todo{make clear that we support only a single relay - add a discussion of extending to multiple relays and what we'd have to do to support latency limitations --> maybe move to a discussion section}

\subsection{Protocol Requirements}

In designing the BRP we had to consider customer requirements,  constraints of the LoRa chip, and FCC regulations.
Beartooth customers want to build networks that cover hundreds of square miles and support voice and data streams.
% Voice transmissions must provide sufficient bandwidth for data streams and deliver packets with at most 500\,ms of latency.
%
The SX1276 LoRa chip, provides 50 10.9\,Kbps channels at SF7, which must carry data as well as control packets.
As such, the protocol must be extremely light-weight and precludes the adoption of existing link-layer protocols.
The FCC limits transmit power to 30\,dbm and each transmission take no more than 400~ms on air without duty-cycle restrictions~\cite{paper:FCC_Title47_P15}.

% We designed a protocol that meets these tradeoffs as follows.
% We limit control packets to include just a few bytes and all data transmissions follow a strict schedule.
% % We limit contention among nodes to very small packets that establish connections to relays -- all other transmissions follow a schedule constructed by a relay.
% Further, we consider the channel capacity within the network and limit the active number of data streams.
% % Further, we configure the network to carry a certain number of voice and binary data streams that fit within channel capacity.
% % Relays broadcast voice streams to all nodes, while binary data streams are addressed to individual nodes, or receiver groups.
% Finally, we restrict each protocol cycle to 500\,ms to meet the latency requirements.

% % We designed our physical layer in accordance to FCC regulations. 
% % With that, we limit our transmit power to 30~dbm and each transmission take no more than 400~ms on air. 
% % Duty-cycle restrictions are not applicable in United States~\cite{paper:FCC_Title47_P15}.

\subsection{Protocol Operation}

\begin{wrapfigure}[17]{r}{0.6\linewidth}
    \vspace{-20pt}
    \centering
    \includegraphics[width=\linewidth]{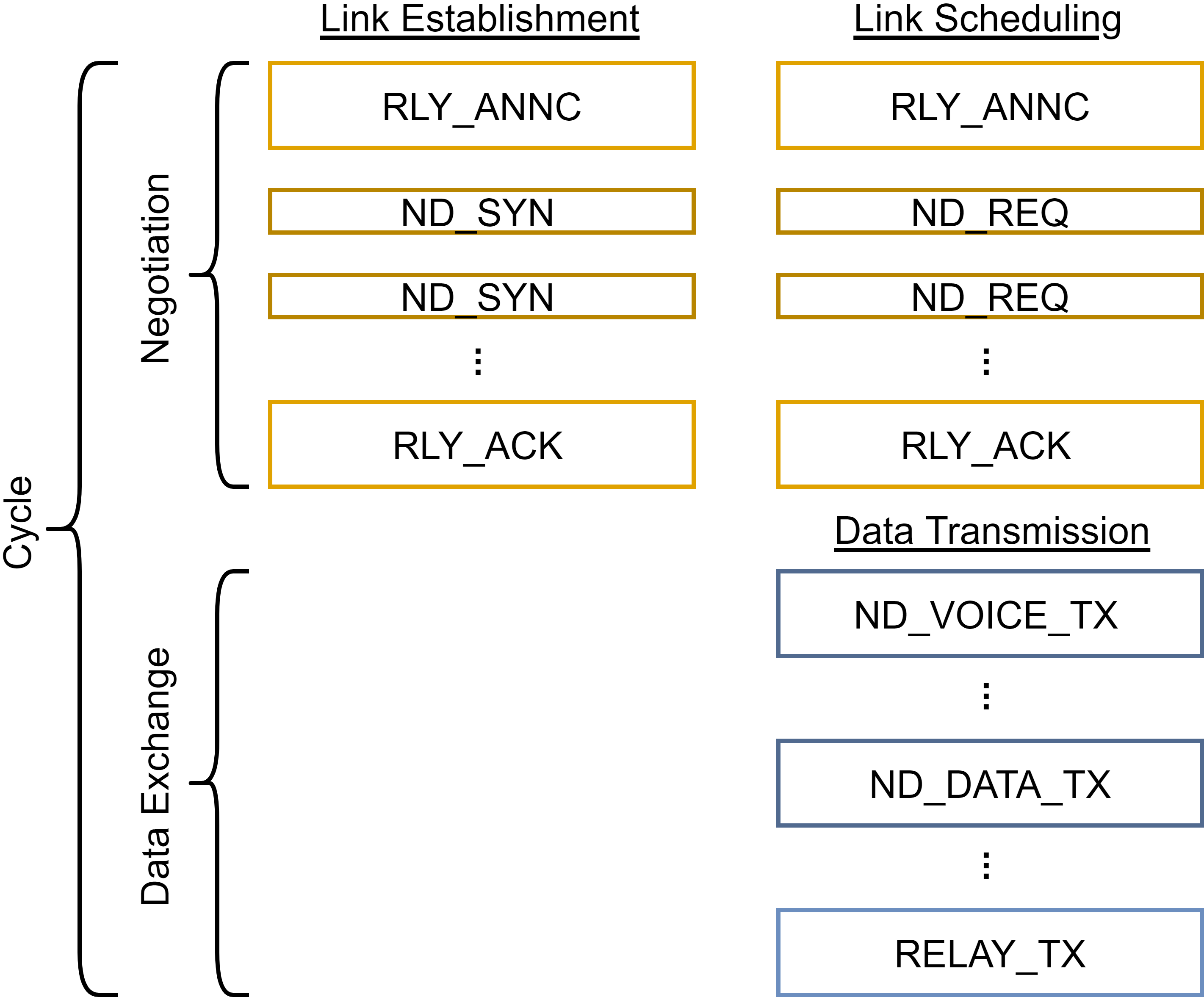}
    \caption{Stages of the Beartooth protocol.}
    \label{fig:protocol_stages}
\end{wrapfigure}

BRP operates in cycles of communication divided into stages, shown in \reffig{fig:protocol_stages}.
During the Negotiation stage nodes either establish a connection, with Link Establishment pattern, or request the relay to schedule transmissions, with the Link Scheduling pattern.
Once a relay schedules node transmissions, the nodes send data to the relay, following the Data Transmission pattern, which then forwards them to other nodes.
While the control frames in Link Establishment may collide, Link Scheduling and Data Transmission messages follow a schedule set by the relay to guarantee collision-free medium access to each user.
% eliminating collisions as well as ensuring high spectral efficiency of the protocol. 
In an event of control frame collision in Link Establishment, relay will not acknowledge connection attempt and nodes will try to establish connection in the next cycle. 
% To support node discovery and mapping between user IDs at the application layer and radio IDs in the hardware, we also implement a version of the address resolution protocol~(ARP).

\subsubsection{Negotiation}

The Negotiation stage establishes links between nodes and the relay and schedules transmissions. 
We currently initialize Beartooth devices as either relays, or nodes, though in future relays will be able to connect to other relays to establish multihop paths in the network.

\paragraph{Link Establishment}

% \todo{mwittie 7/30/19: say that nodes start by establishing a physical link first, including choosing one of the 50 orthogonal channels}

\begin{wrapfigure}{r}{0.5\textwidth}
    \centering
    \vspace{-20pt}
    \includegraphics[scale=0.5]{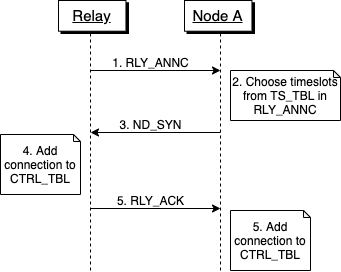}
    \caption{Link Establishment pattern.}
    \label{fig:link_estab}
    \vspace{-20pt}
\end{wrapfigure}

% \begin{figure}[htp]
% \centering
% \subfloat[Link Establishment pattern. \label{fig:link_estab}]{%
%   \includegraphics[width=0.48\textwidth, valign=t]{images/link_establisment.png}%
% }\hfil
% \subfloat[Link Scheduling pattern. \label{fig:link_sched}]{%
%   \includegraphics[width=0.5\textwidth, valign=t]{images/link_scheduling.png}%
% }

% \caption{Figure shows Link Establishment and Link Scheduling patterns.}
% \label{fig:estab_sched}

% \end{figure}

Referring to \reffig{fig:link_estab}, to start a cycle a relay broadcasts the Relay Announce~(\ttt{RLY\_ANNC}) frame (Step 1), which includes the relay hardware ID, configuration ID and a timeslot table~(\ttt{TS\_TBL}) with available timeslots for nodes to establish connections.
The configuration ID specifies required network parameters~(\reftab{tab:parameters} in \refsec{sec:evaluation}).
Nodes use the reception of \ttt{RLY\_ANNC} to synchronize with a relay by establishing the start time of a Cycle. 
In the first iteration of the protocol implementation, we assume there is only one active relay in a network. 
Nodes will stick to a relay and ignore \ttt{RLY\_ANNC} from other relays.
If a node does not receive \ttt{RLY\_ANNC} in some time, then it accepts announcements from other relays.
%This is a precursor to multihop forwarding between relays...
% If there are multiple \ttt{RLY\_ANNC} received by a node from multiple relays, the last announce frame received would be processed and others would be discarded.

To connect with a relay, a node chooses a random available timeslot from \ttt{RLY\_ANNC.TS\_TBL} (2) in which to send the Node Synchronize~(\ttt{ND\_SYN}) frame containing the node ID~(3).
% For example, to send a \ttt{ND\_SYN} in timeslot 3, the node waits for the time it takes to transmit two \ttt{ND\_SYN} frames after the reception of \ttt{RLY\_ANNC}.
The relay collects \ttt{ND\_SYN}s and adds the IDs of connected nodes into its \ttt{CTRL\_TBL}~(4), which will be used to build the \ttt{TS\_TBL} in the next \ttt{RLY\_ANNC}. 
The relay then reflects the IDs of connected nodes in the Relay Acknowledgement~(\ttt{RLY\_ACK}) frame~(5).
A node receiving a \ttt{RLY\_ACK} finds its ID, adds the ID of the relay to its \ttt{CTRL\_TBL}~(6) and considers itself connected on previously acquired timeslot. 
% by adding it to the \ttt{CTRL\_TBL} as well.
If two \ttt{ND\_SYN}s collide, nodes back off and repeat the Link Establishment process.
It is important to note that as \ttt{RLY\_ANNC} advertises only available control timeslots, nodes trying to connect to a relay will only consider available timeslots thus, frames can collide only with other, unconnected nodes. 

\begin{wrapfigure}{r}{0.5\textwidth}
    \centering
    \vspace{-20pt}
    \includegraphics[scale=0.5]{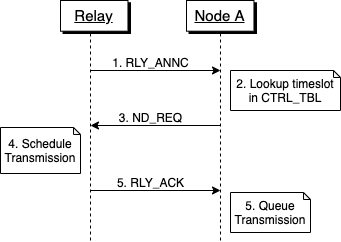}
    \caption{Link Scheduling pattern.}
    \label{fig:link_sched}
    \vspace{-20pt}
\end{wrapfigure}

To ensure that nodes do not need to repeat the connection process, entries in the \ttt{CTRL\_TBL} on both nodes and the relay include a time to live~(TTL) of 10 cycles.
When a node stops receiving \ttt{RLY\_ANNC}, it decrements the TTL of the connection.
Similarly the relay decrements the TTL of a connection, if it does not receive a \ttt{ND\_SYN} or a \ttt{ND\_REQ} (described next) within a cycle.
Reception of these packets resets the TTL to 10.

\paragraph{Link Scheduling}

Referring to \reffig{fig:link_sched}, once a node connects with a relay it may start making requests to schedule its data transmissions.
Upon receiving \ttt{RLY\_ANNC} (1), a node looks up its timeslot in \ttt{CTRL\_TBL}~(2), and sends a Node Request~(\ttt{ND\_REQ}) frame on that timeslot containing a destination address~(3).
In practice, a node may make multiple requests in the same \ttt{ND\_REQ}.

A relay collects all \ttt{ND\_REQ}s and schedules transmissions in the available data slots~(4).
We implement a simple sticky scheduler, which gives priority to continuing flows from the previous cycle up to a limit.
% Otherwise the relay schedules new requests randomly.
We schedule new streams greedily with priority given to voice streams.
If there is no voice stream requested by the node, relay may schedule two data streams instead.
% If nodes do not request transmissions on a particular voice stream, the relay may schedule two binary streams instead.
The stickiness allows nodes to effectively reserve bandwidth for continued data stream, while the limit and randomness provide fairness.

After scheduling, the relay creates a traffic map \ttt{TRFC\_MAP} that assigns nodes with voice and binary data timeslots and sends it out in a \ttt{RLY\_ACK}~(5).
For example, at \ttt{TRFC\_MAP} set to \ttt{[23 45 45 45 67 67]} may mean that node \ttt{23} will use the first timeslot for its binary data transmission, node \ttt{45} will use the next three slots for binary data transmissions, and node \ttt{67} will use the last two slots for a voice transmission on voice stream 3.
Upon receiving \ttt{RLY\_ACK} a node proceeds to queue up its transmissions for the Data Exchange stage~(6).

% In the events of a node found its id in \texttt{traffic\_map}, it quickly extracts the timeslot info from the list and save the timeslot information for the next stage to transmit data on that particular timeslot. 
% If relay don't receive a \texttt{SYN} or \texttt{REQ} packet from a node that is in its \texttt{CtrlTbl}, it will decrease its age by one and eventually, the connection will be deleted from the \texttt{CtrlTbl}.
% Same process applies for node as well. Once a node don't have data to send, it will decrease the age of the connection in its \texttt{CtrlTbl} in every cycle and eventually when it hits zero, it will delete the connection.
% Once the connection is deleted, the connection state of the node will be changed from \texttt{CONNECTED} to \texttt{DISCONNECTED}. 

\vspace{-5pt}

\paragraph{Data Exchange}

\begin{wrapfigure}{r}{0.65\linewidth}
    \vspace{-15pt}
    \centering
    \includegraphics[scale=0.5]{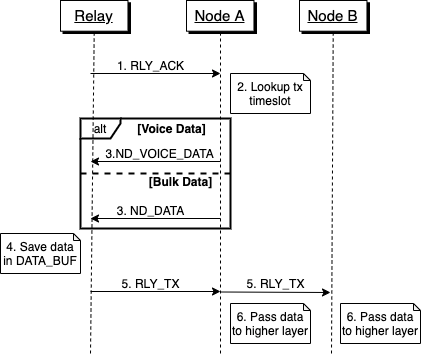}
    \caption{Data Exchange pattern.}
    \label{fig:DATA}
    \vspace{-25pt}
\end{wrapfigure}

Referring to \reffig{fig:DATA}, after a node, here Node A, receives \ttt{RLY\_ACK}~(1) it looks up its transmission timeslot(s) in the \ttt{TRFC\_MAP}~(2).
Depending on the type of transmission it requested in \ttt{ND\_REQ} it forms a Node Voice Transmission~(\ttt{ND\_VOICE\_TX}) frame, or Node Binary Data Transmission~(\ttt{ND\_DATA\_TX}) frames and sends them in the assigned timeslot~(3).
% to avoid collisions with other frames
\ttt{ND\_VOICE\_TX} is addressed implicitly by its timeslot, which allows us to reduce header overhead.
\ttt{ND\_DATA\_TX} contain the address of a particular node, or receiver group as their destination.

The relay collects all \ttt{ND\_VOICE\_TX} and \ttt{ND\_DATA\_TX} frames~(4) and broadcast them in the \ttt{RLY\_TX} frame~(5).
Again \ttt{ND\_VOICE\_TX} are addressed implicitly by their received order.
Nodes receiving the \ttt{RLY\_TX} pass onto the higher layer voice frames if their users subscribe to a particular voice channel, or data packets if \ttt{RLY\_TX} contains \ttt{ND\_DATA\_TX} addressed to them~(6).

% \subsection{Addressing Packets: \textit{Implicit vs Explicit.}}

\paragraph{Implicit Addressing}
\ttt{ND\_VOICE\_TX} frames do not include addressing information to avoid overhead. 
When relay receives them in that particular timeslots, BRP queries \ttt{TRFC\_MAP} to match timeslots with respective \ttt{src\_addr} and \ttt{dst\_addr}.
% Once relay protocol fetches required addresses, protocol proceeds to add them to \ttt{RLY\_TX} frame.

\subsection{Protocol Implementation}

We want to investigate the degree to which it is possible to run a link-layer protocol on the smartphone.
We implement BRP in C++ to be able to run it in hardware, or within an Android app by wrapping it in Kotlin calls.
In our evaluation the Beartooth devices executes physical-layer functions and connects via Bluetooth to the app to process link-layer frames.
As we show later, the Bluetooth interface introduces delays, which force a relaxation of timing with each protocol Cycle.

% \begin{figure}[t]
%     \centering
%     \vspace{-10pt}
%     \includegraphics[scale=0.05]{images/prog_layers.png}
%     \caption{Beartooth Relay Protocol implementation.}
%     \label{fig:prog_layers}
% \end{figure}

\vspace{-10pt}
\section{Evaluation}
\vspace{-5pt}
\label{sec:evaluation}

% \todo{mwittie 3/31/20: discuss how the protocol would work at scale, or in congested environments}
% \todo{How do we know that the Bluetooth link is the limiting factor to performance? justify}
% \todo{compare to different approaches, for example LoRaWAN - can pull data from different papers}
% \todo{the figure for the "two node" scenario (Fig 9) is misleading because it shows
% two relays when in fact the scenario uses only a single relay (based on the
% description)}

\vspace{-20pt}

\begin{figure}[h]
    \centering
    \includegraphics[scale=0.06]{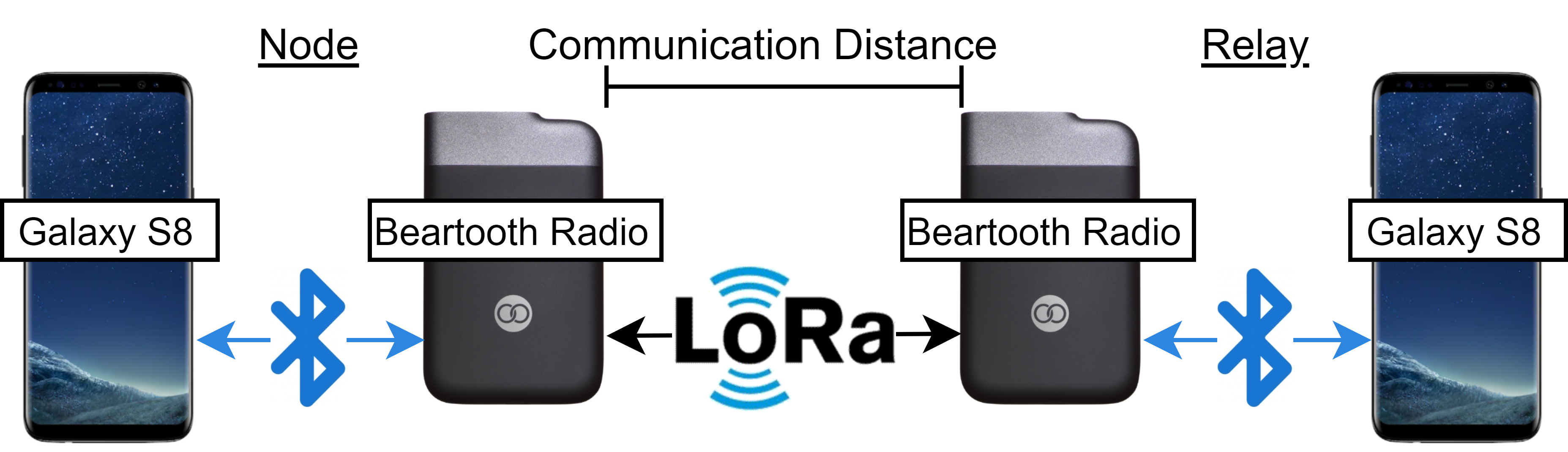}
    \caption{Experimental setup.}
    \label{fig:experimental_setup}
    \vspace{-20pt}
\end{figure}

\begin{wraptable}{r}{0.4\textwidth}
    \small
    \centering
    \vspace{-20pt}
    \begin{tabular}{l|l}
        Parameter & Value \\
        \hline
        \hline
        Center Frequency            & 915\,MHz \\
        Spreading factor            & 7 \\
        Transmit Power              & 30\,dBm\\
        Bandwidth                   & 250\,KHz\\
        Coding Rate                 & 4/5\\
        CRC                         & ON\\
        Voice Encoding Rate         & 1.3\,Kbps\\
        Channel Capacity            & 10.9\,Kbps\\
        Num. Voice Groups           & 2\\
        Num. Data Groups            & 2\\
        % Max Nodes under Relay       & 10 \\
        \hline
        \end{tabular}
    \caption{BRP parameters.}
    \label{tab:parameters}
    \vspace{-20pt}
\end{wraptable}

The goal of the evaluation is demonstrate the feasibility and performance of the software implemented link-layer protocol.
To do so we consider a two hop-scenario illustrated in \reffig{fig:experimental_setup} with one Node connecting to a Relay at some communication distance. 
In our experiments we send the data from a node through a relay back to itself, rather than from one node to another.
Since nodes receive data later in the Cycle it makes no difference if the receiving node is the sender, or another node.
On the other hand, receiving data on the same node allows us to measure end-to-end latency without the need to synchronize clocks.
To be clear, this is for convenience of measurement -- we were able to run all experiments with multiple nodes communicating through a relay as well.
The Beartooth radios connect through Bluetooth to smartphones and connections between nodes use LoRa links.
% which run the protocol on an Android app.
We list LoRa configuration parameters in \reftab{tab:parameters}.

\paragraph{Cycle Duration}

The major challenge to a software implementation of a link-layer protocol is the delay between the physical layer and the software packet processing.
In our implementation the largest and least predictable delay comes from the Bluetooth link between the Beartooth radio and the smartphone.
% The nominal Cycle duration is 500\,ms in the protocol design to accommodate voice latency requirements.
We divide the Data Exchange stage into states of the protocol, during which certain events take place, such as the transmission and reception of particular frame.
If nodes do not follow schedules strictly, the receivers may find themselves in the wrong protocol state and some miss protocol packets.
To accommodate unpredictable delays we increase protocol state and cycle duration by multiplying all timing by a scaling factor.
Our goal in this experiment was to measure the minimum scaling factor and cycle duration in which our protocol can still work correctly.
% the software implementation of our protocol can operate when running in software.
To do so we started the experiment with the scaling factor of 20 and decreased it, until packet collisions and node state mismatch began to cause protocol errors.
% We conducted the experiment in the lab with a negligible distance between nodes of a few feet.

\begin{wrapfigure}{r}{0.5\linewidth}
    \centering
    \vspace{-18pt}
         \centering
         \includegraphics[width=0.5\textwidth]{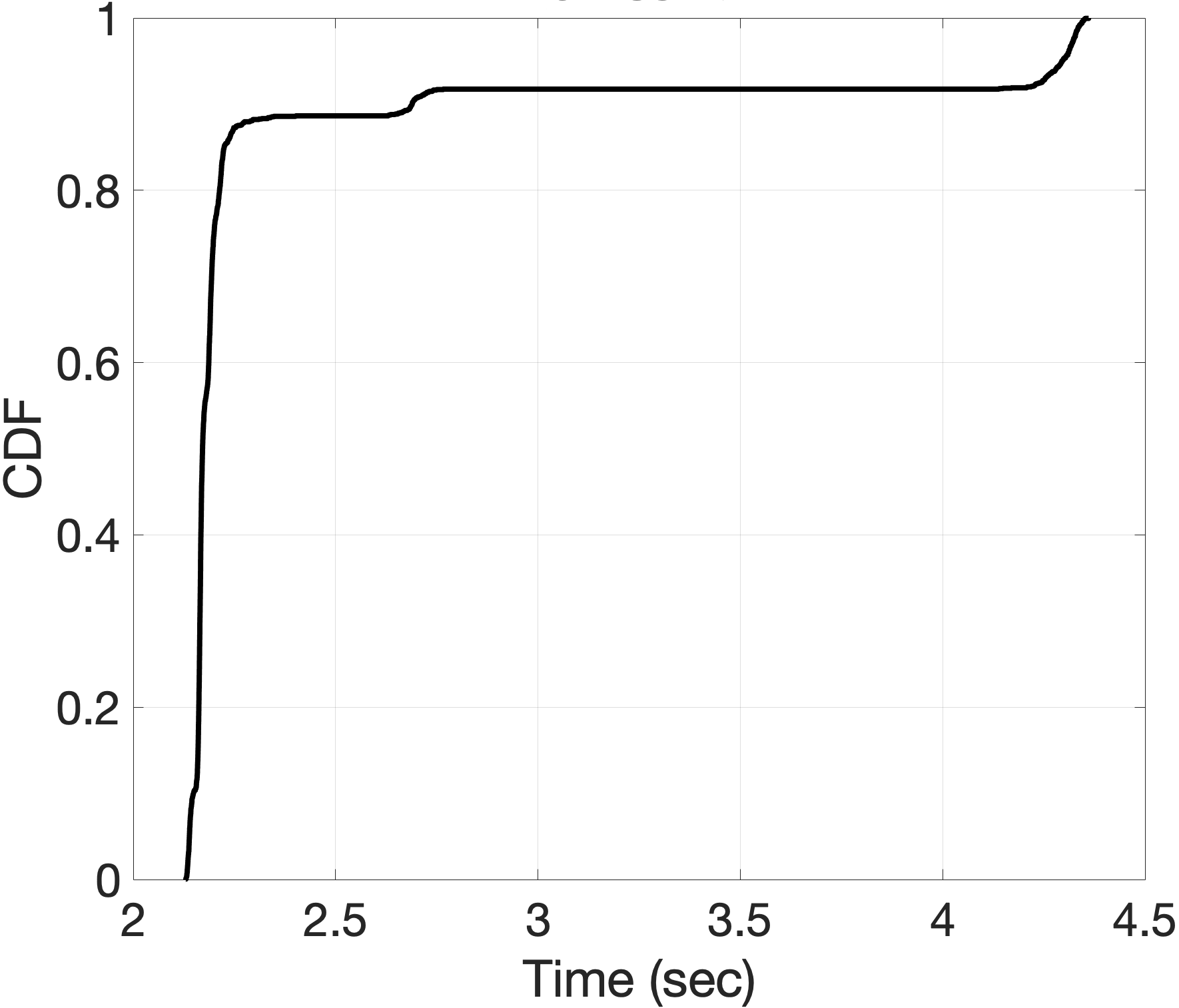}
         \caption{Cycle duration.}
         \label{fig:cycle_dur}
    \vspace{-20pt}
\end{wrapfigure}

\reffig{fig:cycle_dur} shows a CDF of the cycle duration we were able to achieve under the scaling factor of 4.75.
The x-axis shows cycle duration in seconds.
We observe that the average duration of the cycle, measured on the node from the reception of one \ttt{RLY\_ANNC} to another is 2.38\,s.
The result shows that the software implementation of a link-layer protocol is almost 5 times slower than protocol requirements.
This result shows that while software link-layer implementation provides flexibility, the added delay of Bluetooth links makes it impractical in production.
% While the next version of protocol implementation will run on hardware, eliminate Bluetooth delay and decrease cycle duration,  we wanted to understand whether we can expect higher performance in terms of latency and throughput with speed up in cycle duration.
% While the next iteration of Beartooth hardware will run the C++ protocol implementation in hardware to eliminate the Bluetooth delay and decrease cycle duration, we wanted to understand whether we can expect the performance of the protocol in terms of latency and throughput to scale proportionally with the speed up in cycle duration.

\begin{wrapfigure}{r}{0.5\linewidth}
    \centering
    \vspace{-20pt}
         \centering
         \includegraphics[width=0.5\textwidth]{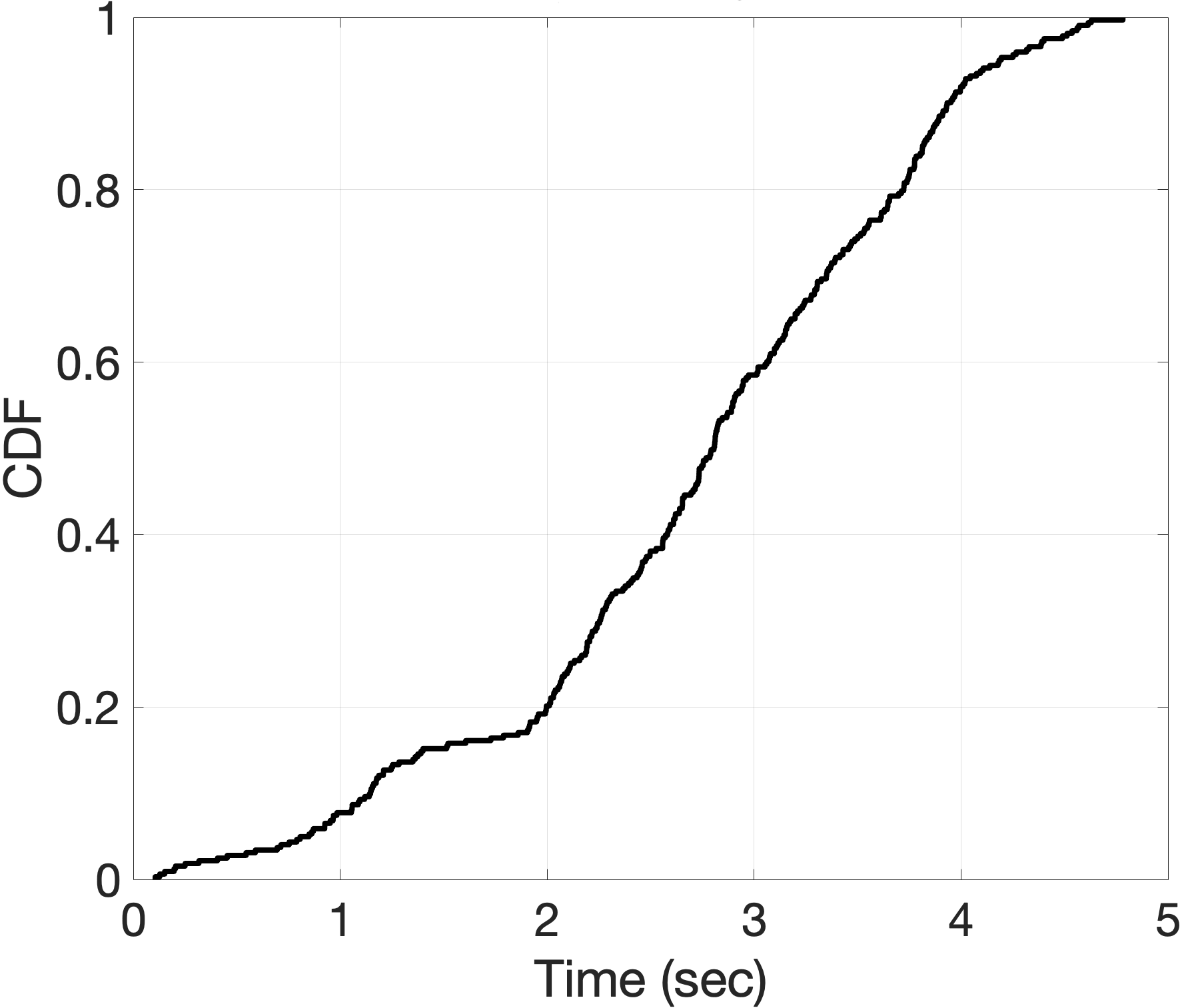}
         \caption{Two-hop latency.}
         \label{fig:latency}
    \vspace{-20pt}
\end{wrapfigure}

\vspace{5pt}
\paragraph{Latency}

% \todo{In the Latency subsection, it is stated that the latency ranged from 0.7 s to 4.78s. According to the graph (Fig 12), the latency had a different range.}

% \begin{figure}
%     \centering
%     \includegraphics[width=\columnwidth]{images/lat_cdf.png}
%     \caption{CDF of latency measurements.}
%     \label{fig:latency}
% \end{figure}

To measure end-to-end latency we bounced back small packets from the node to the relay and back. 
% in \ttt{ND\_DATA\_TX} frames and then back to the node in \ttt{RLY\_TX} frames.
This approach allows us to measure the time difference between transmission and reception on the same node, without the need for synchronizing clocks.
\reffig{fig:latency} shows the CDF of latency measurements.
% We performed the experiment in our lab collecting over 300 measurement samples.
We measure the latency from the point of receiving input application-layer data for transmission to its reception in a \ttt{RLY\_TX}.
We observe that latency ranged from 0.7\,s to 4.78\,s with the mean of 2.71\,s.
The variation in latency comes from the variation of receiving application-layer data in relation to the time in the protocol cycle.
Data received just prior to the transmission of \ttt{ND\_DATA\_TX} achieves the lower latency of less than a full cycle, while data received just after may need to wait for the next cycle to start.
Based on these measurements latency is proportional to cycle duration, and so speeding up cycle duration would proportionally reduce latency. 
% We may, however, need to reduce the nominal cycle duration below 500\,ms to ensure that most packets do not exceed the 500\,ms latency even regardless of their arrival time with respect to protocol state.

% Building upon the latency measurements, minimum RTT  which is the time difference between sending data and receiving an acknowledgement. 
% RTT path is as follows: Node $\rightarrow$ Relay, Relay $\rightarrow$ Node, Node $\rightarrow$ Relay and finally Relay $\rightarrow$ Node.
% Collected data showed RTT is ranging between 0.63 secs to 14.47 secs with an average of 6.26 secs, which translates roughly the twice of the average latency. 
% In Fig.  \ref{fig:lat_cdf} and Fig. \ref{fig:rtt_cdf} latency and RTT CDF functions are shown. 

\paragraph{Throughput}
\label{sec:evaluation:throughput}

To measure application-layer throughput,
% To make sure that we capture any protocol interference that might limit throughput, we transmit data from a sender node, through a relay, to a receiver node different from the sender.
we configure the network to use three voice streams, which equates to six data streams.
We drive the traffic by sending six \ttt{ND\_DATA\_TX} frames in each cycle.
Our results show the average sustained two-hop throughput of 0.782\,Kbps.
This throughput is of course lowered by the scaling factor. 
By moving protocol implementation to hardware, eliminating the Bluetooth link, and lowering scaling factor to 1, we can expect throughput of 3.71\,Kbps.
\ttt{ND\_VOICE\_TX} frames have same time on air as two \ttt{ND\_DATA\_TX}.
Thus, we can expect the throughput of a voice channel to reach 1.3~\,Kbps.
As it is, even with the scaling factor of 4.75 and throughput of 0.782\,Kbps it is possible to transmit voice traffic encoded with Codec2 at 700\,bps~\cite{paper:voice_codec}.
Thus Beartooth radios are the first LoRa system capable of sustained voice transmissions over multiple hops.
We also extrapolate that by reducing the scaling factor to~1, we would be expect to sustain a throughput of 3.71\,Kbps on a 10.9\,Kbps link with the spectral efficiency of 0.341. 
The loss in the spectral efficiency of the protocol comes primarily from scheduling requests in the Negotiation stage. 
% While we cannot increase spectral efficiency by increasing the cycle duration, which would increase latency, the future version of the protocol will make use of less frequent Negotiation opportunities and longer reservations, which will have the same effect of eliminating control traffic overhead.
% Comparing results to commercial satellite BRP multihop protocol implemented on hardware has the potential to exceed the 2.4\,Kbps throughput~\cite{paper:satellite, paper:Sat_cost}.
When multiplied by the 50 orthogonal channels, the 3.71\,Kbps per channel throughput would provide up to 185.5\,Kbps within a deployment area.

% \par
% In the next experiment, we have recorded throughput under different configurations. 
% With current experiment setup, one relay and one node, throughput in the network recorded as 0.184 Kbps. 
% When another node added to the network total bits received by each node increased thus throughput increased, to 0.2366 Kbps. 
% This is because data is segmented into pieces to fit into \texttt{NODE\_DATA\_TX} packet and the maximum size of the packet is defined so that when there are more nodes in the network they also would have chance to send and receive transmissions simultaneously on different timeslots. 
% When this condition is eliminated, maximum throughput the relay network can sustain is measured to be 0.782 Kbps.
% Although, this is not we have aimed to achieve, a few limitations are observed that is later discussed in \refsec{sec:limitations}.

\paragraph{Range}

% \begin{figure}
%     \centering
%     \includegraphics[width=\columnwidth]{images/bt_coverage_range.png}
%     \caption{Map of the range experiment.}
%     \label{fig:range}
% \end{figure}
\begin{wrapfigure}[14]{r}{0.5\linewidth}
\vspace{-20pt}
\centering
\includegraphics[width=0.5\textwidth]{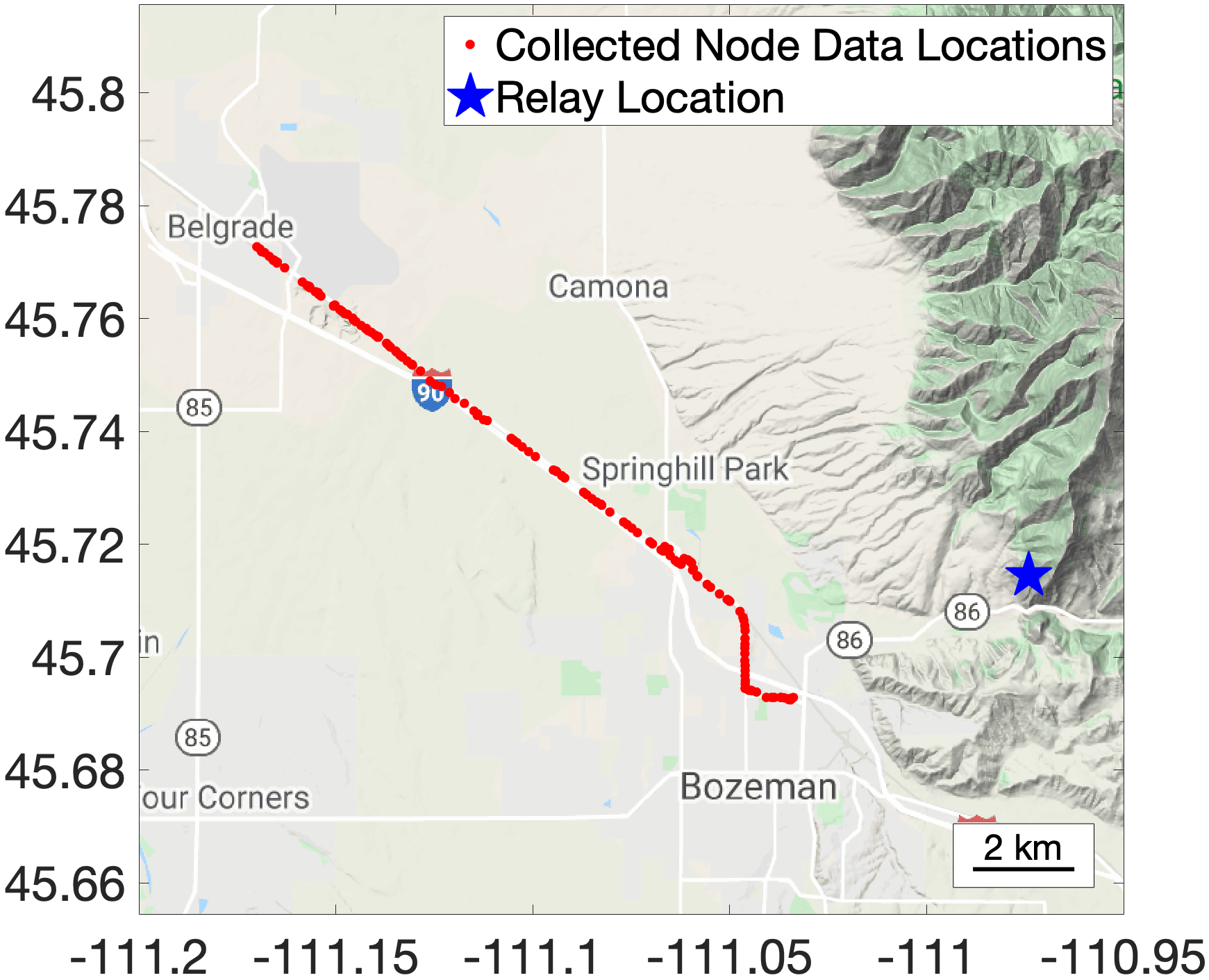}
\vspace{-15pt}
\caption{Transmission range.}
\label{fig:range}
\end{wrapfigure}

We conducted the range experiment with a node sending packets through a relay back to itself.
We place the relay at the height of 250\,m above ground level to create a line of sight link to a vehicle travelling along a roadway, as shown in \reffig{fig:range}.
The x and y-axis show longitude and latitude.
During the experiment we were able to maintain a connection at a distance of 9.5\,mi, or 15.2\,km with the median latency of 2.26\,s corroborating previous results~\cite{paper:LoRaWan_perf, paper:LoRaWan_perf2}.
We note that since the communication is relayed, the coverage area extends to twice the measured distance at 19\,mi, or 30.4\,km.
\looseness-1

\vspace{-10pt}
\section{Limitations and Future Work}
\vspace{-5pt}
\label{sec:lims_and_fw}

While running BRP on the smartphone provides flexibility in development, the approach carries several limitations.

\paragraph{Cycle Duration}
The majority of inflation in cycle time duration comes from the latency of the Bluetooth link between the Beartooth radio and the smartphone.
To eliminate this delay the next generation of Beartooth radio will run BRP in hardware, reducing cycle duration scaling factor to 1.
We will also reduce signalling within the protocol by making the Negotiation stage less frequent and reservations longer-lived by spanning multiple Data Exchange stages.

% We plan on reducing that delay by experimenting with a higher-rate WiFi direct link between the Beartooth radio and the smartphone and by eliminating that link altogether on relay nodes by moving BRP onto the radio hardware.
% By being able to reduce the cycle duration scaling factor, we will improve network latency and throughput.

% We also concluded that cycle duration has to be shorter to meet the 500\,ms latency deadlines for voice flows.
% Accordingly, we will shorten the cycle and reduce signalling within the protocol by making the Negotiation stage less frequent and reservations longer-lived by spanning multiple Data Exchange stages.
% We will also redesign the Negotiation stage to allow nodes to share control timeslots, which will reduce control traffic and increase the maximum number of connections to a relay.

\paragraph{Multihop Paths}

The current implementation of BRP supports only two-hop paths. 
However, we designed BRP with the intent of extending it to multihop paths, by forwarding data between relays.
We believe that scheduled transmissions within BRP will be helpful in ensuring low latency and sufficient throughput for voice streams, but functionality around routing and bandwidth reservation will need novel, light-weight solutions to work reliably on LoRa's low-bandwidth links.
% Our approach will focus on a distributed transmission scheduling solution at the link-layer and adopt an existing routing protocol.
% We are also in the process of implementing a version of the Address Resolution Protocol~(ARP) to map between user IDs (eventually IP addresses) and radio IDs in the hardware to support node discovery and routing protocol operation.

% \paragraph{Non Line-of-sight Paths}

% While outside of the scope of this paper, we will also explore the performance of the Beartooth radio under various radio settings and in different topological scenarios, including non line-of-sight~(LOS) transmissions.
% The existing work on LoRa measurement assumes LOS scenarios and so we replicated those in this paper~\cite{paper:LoRaWan_perf, paper:LoRaWan_perf2}, for example by collecting the results in \refsec{sec:evaluation:throughput} with the relay node placed at elevation.
% We are interested, however, in quantifying the extent to which multihop connectivity can mitigate shadowing of LoRa transmissions and continue to deliver real-time application traffic.

\paragraph{Mobile Application Traffic}

Finally, we would like to experiment with real mobile application traffic beyond voice streams.
The flexibility of BRP implementation means that the protocol may be reconfigured to match channel characteristics to message sizes as well as bandwidth and latency requirements of individual applications.
We would like to evaluate whether a link-layer protocol can dynamically adapt to application layer requirements in a resource constrained network.
We would also like to explore the possibility of modifying current mobile app functionality to take advantage of local area network connectivity, when a link to the cloud is not available.

\vspace{-10pt}
\section{Conclusion}    
\vspace{-5pt}
\label{sec:conclusions}

Infrastructure-based networks based on cellular and WiFi coverage struggle to provide full coverage.
In this paper we position LoRa-based Beartooth networks as a third leg of communications capable of supporting mobile application traffic beyond cellular and WiFi networks.
We present an experimental relaying protocol for Beartooth radios and evaluate the performance of its software-based implementation.
While ultimately the protocol will be deployed on Beartooth radios in hardware, even when running on a smartphone it provides sufficient throughput for voice flows, making Beartooth the first LoRa system to do so.
This study will inform future iterations of the Beartooth Relay Protocol to support offline mobile application traffic.

% the Beartooth radio paired with a software implementation of a relaying

% In this paper we present a new LoRa-based radio and an experimental relaying protocol designed multihop link-layer protocol intended to 

% Existing infrastructure-based connectivity solutions such as cellular networks and WiFi predominantly and widely available to user, they come with set of limitations. 
% Although some limitations can be addressed with different technologies like satellite communications, a tradeoff has to be made regarding performance and cost. 
% In this paper we have introduced a multihop, long range, relatively more battery efficient solution that utilizes LoRa modulation. 
% We mentioned how resource allocation and management has been done along with different stages of the protocol. 
% And finally we have presented evaluations of the protocol by presenting the key parameters such as latency, throughput, battery consumption and range.
% We believe Beartooth Relay Protocol can be used as a third leg and a complementary technology where limitations of the infrastructure-based solutions surfaces.

%  \section{Acknowledgement}
 
% We would like to thank our colleagues Seraj Al Mahmud, Reha Abbasi and Berrak Erturk for their help and assistance with the range measurements.

\vspace{-10pt}

\footnotesize
\bibliographystyle{./bibliography/splncs03_unsrt}
\bibliography{./bibliography/paper}

\end{document}